\documentclass{article}
\usepackage{amsmath}
\usepackage{times}
\usepackage[dvips]{graphicx}
\usepackage{color}
\usepackage{mathtools}
\usepackage{graphicx}
\usepackage{amsmath,amssymb}
\usepackage{mathtools}
\def\spcc{{ \mbox{\hspace{.1in}}}}

\newcommand{\ld}{\lambda}
\newcommand{\frd}[2]{{\displaystyle \frac{#1}{#2}}}

\newcommand{\be}{\begin{eqnarray}}
\newcommand{\ee}{\end{eqnarray}}
\newcommand{\ben}{\begin{eqnarray*}}
\newcommand{\een}{\end{eqnarray*}}
\newcommand{\nn}{\nonumber}

\newcommand{\tg}{\mbox{\boldmath $g$}}

\newcommand{\tu}{\mbox{\boldmath $u$}}

\newcommand{\tI}{\mbox{\boldmath $I$}}

\newcommand{\tQ}{\mbox{\boldmath $Q$}}

\newcommand{\tr}{\mbox{tr}}

\def\rr#1{(\ref{#1})} 

\newcommand{\ar}[1]{{\bf{A}}^{{(#1)}}}
\newcommand{\vr}[1]{{\bf{v}}^{{(#1)}}}
\newcommand{\cba}[1]{\bar{A}^{{(#1)}}}
\newcommand{\cbv}[1]{\bar{v}^{{(#1)}}}
\newcommand{\cbvs}[1]{\bar{v}^{{(#1)2}}}
\newcommand{\ca}[1]{A^{{(#1)}}}
\newcommand{\cv}[1]{v^{{(#1)}}}
\newcommand{\cvs}[1]{v^{{(#1)2}}}
\newcommand{\cas}[1]{A^{{(#1)2}}}
\newcommand{\Iv}[2]{I^{{(#1,#2)}}}
\newcommand{\ars}[2]{{\bf{A}}^{{(#1)#2}}}

\def \tr{\mbox{tr\hskip 1pt}}

\def\theequation{\arabic{equation}}
\addtocounter{figure}{0}

\begin{document}

\title{The number of independent invariants for m unit vectors and n symmetric second order tensors
is 2m+ 6n-3}


\author{MHBM Shariff \\
              Department of Mathematics,
 Khalifa University, UAE.}
\date{}
\maketitle
\begin{abstract}

Anisotropic invariants play an important role in continuum mechanics. Knowing the number of independent invariants is crucial in modelling  and in a rigorous construction
of a constitutive equation for a particular material, where it is determined by doing
tests that hold all, except one, of the independent invariants constant so that the
dependence in the one invariant can be identified. Hence, the aim of this paper is to prove that the number of independent invariants for a set of $n$ symmetric tensors and $m$ unit vectors is at most $2m+ 6n-3$. We also give relations between classical invariants in the corresponding minimal integrity
basis.

\end{abstract}

\section{Introduction}

The construction of rotational invariants of sets containing vectors and tensors in continuum mechanics, especially for anisotropic materials
\cite{ahamed16,shabusmer17,shariffetal18,soldatos18,soldatos19}, has been active for around $70$ years. The "classical" invariants and minimal integrity bases constructed in Spencer \cite{spencer71}, published in 1971,
have been extensively used in the literature.
Spencer \cite{spencer71} stated that " It
{\it frequently} happens that polynomial relations exist between invariants which do not permit
any one invariant to be expressed as a polynomial in the remainder. Such relations are called
syzygies". This suggests that some of the invariants in a given minimal integrity basis may
not be independent (see also \cite{miller2004} ). However, due to
the difficulty in constructing relations (syzygies) among classical invariants, Spencer \cite{spencer71} did not specifically mention the number
of existing syzygies for a given minimal integrity basis, and in view of this, the number
of independent invariants was, often, not correctly stated in the literature and it is occasionally assumed in the literature (see, for example
\cite{ahamed16,holzapfel09a} ) that all the invariants 
in a minimal integrity  (or irreducible functional \cite{boehler77}) basis are independent.
To prove the number of independent classical invariants in a minimal integrity (or irreducible functional) basis is not straightforward. However, recently, in the case of an m-preferred direction anisotropic solid, Shariff \cite{shariff16} 
proved that the number of independent invariants is at most $2m+3$ (see also \cite{sharbusta15,shariff13,rubin15,aguiar17,aguiar18}) and in the case of $n-symmetric$ tensors, Shariff
\cite{shariff18} (see also \cite{aguiar18a}) proved that at most $6n-3$ classical invariants are independent. In this communication, we extend these result to prove that the number of independent
invariants for $m$ vectors and $n$ symmetric tensors is $2m+ 6n-3$. We also give relations (not necessarily syzygies) between classical invariants in the corresponding minimal integrity
basis. Knowing the number of independent invariants is crucial in modelling \cite{soldatos18,soldatos19} and in a rigorous construction
of a constitutive equation of a particular material, where it is determined by doing
tests that hold all, except one, of the independent invariants constant so that the
dependence in the one invariant can be identified \cite{holzapfel09,humphreyetal1990,shariff2008}.

\section{Preliminaries}\label{sec-1}
In this paper, the summation convention is not used and  all subscripts $i$,$j$ and $k$ take the values $1,2,3$, unless stated otherwise.
Consider the set 
\be\label{pr1} S(m,n) = \{ \vr{r}, \ar{s} \mid r=0,1 \ldots , m \, , s=1,2 \ldots , n \} \, , \ee
where $m$ and $n$ are non-negative integers, $\ar{s}$ are symmetric tensors defined on a three-dimensional Euclidean space and
$\vr{r}$ are linearly independent unit vectors. Using a fixed Cartesian basis $\{ \tg_1,\tg_2,\tg_3 \}$, we have,
\be\label{pr2}
\vr{r} =\sum_{i=1}^3 \cbv{r}_i \tg_i \, , \spcc \ar{s}=\sum_{i,j=1}^3 \cba{s}_{ij} \tg_i\otimes\tg_j \, , \ee
where $\otimes$ denotes the dyadic product and since $\ar{s}$ is symmetric $\cba{s}_{ij} =\cba{s}_{ji}$. Since
$\vr{r}$ are unit vectors, we have,
\be\label{pr3}
\sum_{i=1}^3 \cbvs{r}_i = 1 \, , \spcc r=1,2 \ldots , m \, . \ee
In view of \rr{pr2} and \rr{pr3} we can say that at most there are $2m + 6n$ independent component variables in \rr{pr1}.

Since the classical invariants in a minimal integrity basis are traces of tensors and dot products of vectors \cite{spencer71} and they are explicit functions of the $2m+6n$ components, hence
the number of independent invariants in a minimal integrity basis for the set $S(m,n)$ cannot be greater than $2m+6n$.
It is important to note that the components $\cbv{r}$ and $\cba{s}_{ij}$ are {\it not} invariants and hence they cannot be explicitly expressed
as a function of the classical invariants in a minimal integrity basis. However if we use the spectral basis $\{ \tu_1,\tu_2,\tu_3 \}$ where its elements
are eigenvectors of $\ar{1}$, we have
\be\label{pr4}
\ar{1} &=& \sum_{i=1}^3 \ld_i \tu_i\otimes\tu_i \, , \spcc \vr{r} = \sum_{i=1}^3 \cv{r}_i\tu_i \, , \spcc
\ar{s} = \sum_{i,j=1}^3 \ca{s}_{ij} \tu_i\otimes\tu_j \, , \nn \\
\spcc r&=&1,2, \ldots , m \, , s=2,3, \ldots , n \, . \ee
In this case, the spectral components $\cv{r}_i \, ,\ca{s}_{ij}$ are rotational{\it invariants} with respect to the rotation tensor $\tQ$, since
\[ \cv{r}_i  = \vr{r}\cdot\tu_i = \tQ\vr{r}\cdot\tQ\tu_i \, ,\spcc \ca{s}_{ij} = \tu_i\cdot \ar{s}\tu_j = \tQ\tu_i\cdot \tQ\ar{s}\tQ^T\tQ\tu_j \, . \]
\be\label{pr5} \ee
Hence, it is possible that the spectral components can be expressed explicitly in terms of the classical invariants in a given minimal integrity basis.
Since $\vr{r}$ are unit vectors,
\be\label{pr3a}
\sum_{i=1}^3 \cvs{r}_i  = 1 \, , \spcc r=1,2 \ldots , m \, . \ee

It is clear that the number of independent spectral invariants cannot be greater $2m + 6n-3$ and hence the number of independent classical
invariants  in a $S(m,n)$-minimal integrity basis is at most $2m + 6n-3 < 2m + 6n$

In the Section \ref{sec-2}, based on the work of Shariff \cite{shariff16,shabusmer17,shariff18} and Aguiar and Rocha \cite{aguiar18a} , we show relations between the classical invariants in a
$S(m,n)$-minimal integrity basis using our spectral invariants. We assume, for simplicity, $\ar{1}$ is invertible and its eigenvalues $\ld_i$ are distinct. 
In the case when the eigenvalues $\ld_i$ and some of the eigenvalues of $\ar{s}$ are not distinct, the number of independent invariants is far
less than $2m + 6n-3$, as exemplified in the Appendix A.

In the case of $S(0,n)$ it is shown in references \cite{shariff18,aguiar18a} that the number of independent invariants is $6n-3$ and, in these references, classical invariant relations are given.

\section{Classical invariant relations}\label{sec-2}
In this section, we first obtain relations for certain values of $m$ and $n$ and then derive relations for general $m$ and $n$.
The construction of these relations require the invariants
\be\label{ci-1} I_1&=&\tr \ar{1} = \sum_{i=1}^3 \ld_i \, , \nn \\
 I_2&=& \frd{1}{2}\left((\tr\ar{1})^2-\tr\ars{1}{2}\right) = \ld_1\ld_2+\ld_1\ld_3+\ld_2\ld_3\, , 
\nn \\
I_3&=& \det(\ar{1}) =\ld_1\ld_2\ld_3 \,  . \ee
It is commonly known that the above relations are independent and since the eigenvalues $\ld_i$ are independent, there are no relations between the classical invariants and hence
the three classical invariants are independent. The eigenvalues $\ld_i$ we can be explicitly expressed in terms of the classical invariants \cite{itskovbook} , i.e.,
\be\label{eig1} \ld_i = \frd{1}{3} \left\{ I_1 + 2\sqrt{I_1^2 - 3I_2} \cos\frd{1}{3}[\theta + 2\pi(i-1)] \right\} \, , \spcc i=1,2,3 \, ,
 \ee
where
\be\label{eig2} \theta = \arccos \left[ \frd{2(I_1^3 - 9I_1I_2 + 27I_3}{2[I_1^2-3I_2]^\frac{3}{2}}\right]  \, , \ee
taking note that since th eigenvalues $\ld_i$ are distinct, we have, $I_1^2-3I_2 \ne 0$.

\subsection{$m=1, n=1$. Only $5$ classical invariants are independent}
The set $S(1,1)$ is generally associated with transversely isotropic elastic materials, where their strain energy functions can be written in the form
\be\label{ci-2}
W(\vr{1}\otimes\vr{1},\ar{1}) = \hat{W}(\vr{1},\ar{1}) \, . \ee
The $5$ invariants in the minimal intetigrity basis are
\be\label{ci-3} \Iv{1}{1}_1&=& I_1 \, , \spcc 
 \Iv{1}{1}_2 = I_2 \, , \spcc \Iv{1}{1}_3=  I_3 \, , \nn \\
\Iv{1}{1}_4&=&\vr{1}\cdot\ar{1}\vr{1} = \sum_{i=1}^3 \cvs{1}_i\ld_i \, , \nn \\
\Iv{1}{1}_5 &=& \vr{1}\cdot\ars{1}{2}\vr{1} = \sum_{i=1}^3 \cvs{1}_i\ld_i^2 \, . \ee
In view of \rr{pr3a}, only $5$ of the spectral invariants are independent and since there are no relations between the $5$ spectral
invariants,
its clear that there are no relations between the classical invariants. Hence, we have $5$ independent classical invariants.

\subsection{$m=2, n=1$.  At most $7$ classical invariants are independent}
The set $S(2,1)$ is commonly associated with the strain energy function of an elastic solid with two preferred directions, i.e.
\be\label{ci-4} W(\vr{1}\otimes\vr{1},\vr{2}\otimes\vr{2},\ar{1}) = \hat{W}(\vr{1},\vr{2},\ar{1}) \, . \ee
There are ten classical invariants in the minimal integrity basis, i.e.,
\be\label{ci-5} \Iv{2}{1}_1&=& I_1 \, , \spcc 
 \Iv{2}{1}_2 = I_2 \, , \spcc \Iv{2}{1}_3=  I_3 \, , \\
\label{ci-5a} \Iv{2}{1}_4 &=& \Iv{1}{1}_4 =\sum_{i=1}^3 \cvs{1}_i\ld_i\, , \spcc \Iv{2}{1}_5 =   \Iv{1}{1}_5 =\sum_{i=1}^3 \cvs{1}_i\ld_i^2\, ,   \\
\label{ci-5b} \Iv{2}{1}_6 &=&\vr{2}\cdot\ar{1}\vr{2} = \sum_{i=1}^3 \cvs{2}_i\ld_i \, , \spcc 
\Iv{2}{1}_7 = \vr{2}\cdot(\ar{1})^2\vr{2} = \sum_{i=1}^3 \cvs{2}_i\ld_i^2 \, , \nn \\
\label{ci-5bb} &&  \\
\Iv{2}{1}_8 &=& (\vr{1}\cdot\vr{2})^2 = (\sum_{i=1}^3 \cv{1}_i\cv{2}_i)^2 \, , \nn \\
\Iv{2}{1}_9 &=& (\vr{1}\cdot\vr{2})\vr{1}\cdot\ar{1}\vr{2} = (\sum_{i=1}^3 \cv{1}_i\cv{2}_i)\sum_{i=1}^3 \cv{1}_i\cv{2}_i \ld_i \nn \\
\label{ci-5c} \Iv{2}{1}_{10} &=& (\vr{1}\cdot\vr{2})\vr{1}\cdot(\ar{1})^2\vr{2} = (\sum_{i=1}^3 \cv{1}_i\cv{2}_i)\sum_{i=1}^3 \cv{1}_i\cv{2}_i \ld_i^2  \, . \ee

Shariff and Bustamante \cite{sharbusta15} have given $3$ classical invariant relations and hence only $7$ of the above invariants
are independent. However, below we give alternative invariant relations to strengthen our claim that only $7$ of the classical invariants
are independent.

From \rr{eig1} and \rr{ci-5}, $\ld_i$ is expressed in terms of $\Iv{2}{1}_i$. From \rr{pr3a}, with $r=1,2$,  \rr{ci-5a} and \rr{ci-5b} we have
$6$ linear equations in $\cvs{1}_i$ and $\cvs{2}_i$. On solving the two $3$ linear equations (see Appendix B) in turn we can express $\cvs{1}_i$ and $\cvs{2}_i$ explicitly in terms
of $\Iv{2}{1}_{\alpha} \, , \alpha=1,2 \ldots , 7$. Taking into consideration the sign of $\cv{1}$ and $\cv{2}$ and the fact that  $\Iv{2}{1}_8, \Iv{2}{1}_9 ,
\Iv{2}{1}_{10}$ can be explicitly expressed in terms of $\ld_i, \cv{1}, \cv{2}$, it is clear that, in view of \rr{ci-5c}, they can be expressed explicitly in terms of $\Iv{1}{2}_{\alpha} \, , \alpha=1,2 \ldots , 7$. Hence only $7$ classical invariants are independent.

\subsection{$m=1, n=2$. At most $11$ invariants are independent}\label{sec-m1n2}
An example of a $S(1,2)$ constitutive function of the form
\be\label{ci-6} 
W(\vr{1}\otimes\vr{1},\ar{1},\ar{2}) = \hat{W}(\vr{1},\ar{1},\ar{2}) \,    \ee
can be found in Shariff et. al. \cite{shariffetal18}.
There are $18$ classical invariants in the minimal integrity basis for the function \rr{ci-6}, i.e.,
\be\label{ci-6a} \Iv{1}{2}_1&=& I_1 \, , \spcc 
 \Iv{1}{2}_2 = I_2 \, , \spcc \Iv{1}{2}_3=  I_3 \, , \\
\Iv{1}{2}_4 &=&\vr{1}\cdot\ar{1}\vr{1} = \sum_{i=1}^3 \cvs{1}_i\ld_i \, , \spcc
\label{ci-6b} \Iv{1}{2}_5 = \vr{1}\cdot\ars{1}{2}\vr{1} = \sum_{i=1}^3 \cvs{1}_i\ld_i^2 \,  , \nn \\
&& \\
\label{ci-6c}
\Iv{1}{2}_6 &=& \tr\ar{2} = \sum_{i=1}^3 \ca{2}_{ii} \, , \spcc  \Iv{1}{2}_7 \tr(\ar{2}\ar{1}) = \sum_{i=1}^3 \ld_i \ca{2}_{ii} \, , \nn \\
 \Iv{1}{2}_8 &=& \tr(\ar{2}\ars{1}{2}) = \sum_{i=1}^3 \ld_i^2\ca{2}_{ii}  \, , \ee
\be\label{ci-6d} \Iv{1}{2}_9 &=& \tr(\ars{2}{2}) =  \cas{2}_{11} + \cas{2}_{22} + \cas{2}_{33} + 2(\cas{2}_{12} + \cas{2}_{13} + \cas{2}_{23}) \, ,  \nn \\
\Iv{1}{2}_{10} &=& \tr(\ars{2}{2}\ar{1}) =\ld_1(\cas{2}_{11} + \cas{2}_{12} + \cas{2}_{13}) + \nn \\ 
&& \ld_2(\cas{2}_{21} + \cas{2}_{22} + \cas{2}_{23}) + \ld_3(\cas{2}_{31}+ \cas{2}_{32} + \cas{2}_{33}) \, , \nn \\
\Iv{1}{2}_{11} &=& \tr(\ars{2}{2}\ars{1}{2}) = \ld_1^2(\cas{2}_{11} + \cas{2}_{12}+ \cas{2}_{13}) +  \nn \\
&& \ld_2^2(\cas{2}_{21} + \cas{2}_{22} + \cas{2}_{23}) + \ld_3^2(\cas{2}_{31} + \cas{2}_{32} + \cas{2}_{33}) \, , \ee
\be\label{ci-6e}
\Iv{1}{2}_{12} &=& \vr{1}\cdot\ar{2}\vr{1} \, , \spcc
\Iv{1}{2}_{13} = \vr{1}\cdot\ars{2}{2}\vr{1} \, , \spcc \Iv{1}{2}_{14}  = \tr\ars{2}{3} \, , \nn \\
\Iv{1}{2}_{15}  &=& \vr{1}\cdot(\ar{1}\ar{2}\vr{1}) \, , \spcc \Iv{1}{2}_{16}  = \vr{1}\cdot(\ar{1}\ars{2}{2}\vr{1}) \, , \nn \\
\Iv{1}{2}_{17}  &=& \vr{1}\cdot(\ar{2}\ars{1}{2}\vr{1}) \, , \spcc \Iv{1}{2}_{18}  =\vr{1}\cdot(\ars{1}{2}\ars{2}{2}\vr{1}) \, . \ee

We note that the invariants $\Iv{1}{2}_{\alpha} \, , \alpha=12,13,\ldots , 18$ can be explicitly expressed in terms of $\ld_i, \cv{1}_i$ and
$\ca{2}_{ij}$ but, for brevity, we omit such explicit expressions.

From \rr{eig1} and \rr{ci-6a}, $\ld_i$ is expressed in terms of $\Iv{1}{2}_i$. From \rr{pr3a}, with $r=1$, and \rr{ci-6b}, we have
$3$ linear equations in $\cvs{1}_i$.  On solving these linear equations we can express $\cvs{1}_i$ explicitly in terms
of $\Iv{1}{2}_{\alpha} \, , \alpha=1,2 \ldots , 5$.  The invariants $\ca{2}_{ii}$ can be expressed in terms of $\Iv{1}{2}_{\alpha} \, , 
\alpha=1,2,3,6,7,8$ by solving the $3$ linear equations in \rr{ci-6c} for $\ca{2}_{ii}$. In Eqn. \rr{ci-6d}, the invariants $\cas{2}_{12}, \cas{2}_{13}, \cas{2}_{23}$ appear
linearly. Hence we can solve the $3$ linear equations so that these invariants (see Appendix B) can be expressed in terms of $\Iv{1}{2}_{\alpha} \, , \alpha=1,2,3, 6,7 \ldots , 11$.
Since the classical invariants $\Iv{1}{2}_{\alpha} \, , \alpha=12, 13 , \ldots, 17$ can be explicitly expressed in terms of  $\ld_i, \cv{1}_i$,
$\ca{2}_{ij}$, and taking the appropriate sign for $\cv{1}_i$ and $\ca{2}_{ij}$, they can be expressed in terms of $\Iv{1}{2}_{\alpha} \, , \alpha=1,2, \ldots, 11$,
indicating that only $11$ classical invariants are independent.

\subsection{$m=2, n=2$. At most $13$ invariants are independent}
An example of a $S(2,2)$ constitutive function of the form
\be\label{ci-7} 
W(\vr{1}\otimes\vr{1},\vr{2}\otimes\vr{2},\ar{1},\ar{2}) = \hat{W}(\vr{1},\vr{2}, \ar{1},\ar{2}) \,   \ee
can be found in \cite{shariffetal19}.
There are $37$ classical invariants for the function  \rr{ci-7} in the minimal integrity basis and they are:
\be\label{ci-7a} \Iv{2}{2}_1&=& I_1 \, , \spcc 
 \Iv{2}{2}_2 = I_2 \, , \spcc \Iv{2}{2}_3=  I_3 \, , \\
\Iv{2}{2}_4 &=&\vr{1}\cdot\ar{1}\vr{1} = \sum_{i=1}^3 \cvs{1}_i\ld_i \, , \spcc
\label{ci-7b} \Iv{2}{2}_5 = \vr{1}\cdot\ars{1}{2}\vr{1} = \sum_{i=1}^3 \cvs{1}_i\ld_i^2 \,  , \nn \\
\Iv{2}{2}_6 &=&\vr{2}\cdot\ar{1}\vr{2} = \sum_{i=1}^3 \cvs{2}_i\ld_i \, , \spcc 
\Iv{2}{2}_7 = \vr{2}\cdot\ars{1}{2}\vr{2} = \sum_{i=1}^3 \cvs{2}_i\ld_i^2 \, , \nn \\
&& \\
\label{ci-7c}
\Iv{2}{2}_8 &=& \tr\ar{2} = \sum_{i=1}^3 \ca{2}_{ii} \, , \spcc  \Iv{2}{2}_9 \tr(\ar{2}\ar{1}) = \sum_{i=1}^3 \ld_i \ca{2}_{ii} \, , \nn \\
 \Iv{2}{2}_{10} &=& \tr(\ar{2}\ars{1}{2}) = \sum_{i=1}^3 \ld_i^2\ca{2}_{ii}  \, , \ee
\be\label{ci-7d} \Iv{2}{2}_{11} &=& \tr(\ars{2}{2}) =  \cas{2}_{11} + \cas{2}_{22} + \cas{2}_{33} + 2(\cas{2}_{12} + \cas{2}_{13} + \cas{2}_{23}) \, ,  \nn \\
\Iv{2}{2}_{12} &=& \tr(\ars{2}{2}\ar{1}) =\ld_1(\cas{2}_{11} + \cas{2}_{12} + \cas{2}_{13}) + \nn \\ 
&& \ld_2(\cas{2}_{21} + \cas{2}_{22} + \cas{2}_{23}) + \ld_3(\cas{2}_{31}+ \cas{2}_{32} + \cas{2}_{33}) \, , \nn \\
\Iv{2}{2}_{13} &=& \tr(\ars{2}{2}\ars{1}{2}) = \ld_1^2(\cas{2}_{11} + \cas{2}_{12}+ \cas{2}_{13}) +  \nn \\
&& \ld_2^2(\cas{2}_{21} + \cas{2}_{22} + \cas{2}_{23}) + \ld_3^2(\cas{2}_{31} + \cas{2}_{32} + \cas{2}_{33}) \, , \ee
\be\label{ic-7e}
\Iv{2}{2}_{14} &=& \vr{1}\cdot\ar{2}\vr{1} \, , \spcc
\Iv{2}{2}_{15} = \vr{1}\cdot\ars{2}{2}\vr{1} \, , \spcc \Iv{2}{2}_{16}  = \tr\ars{2}{3} \, , \nn \\
\Iv{2}{2}_{17}  &=& \vr{1}\cdot(\ar{1}\ar{2}\vr{1}) \, , \spcc \Iv{2}{2}_{18}  = \vr{1}\cdot(\ar{1}\ars{2}{2}\vr{1}) \, , \nn \\
\Iv{2}{2}_{19}  &=& \vr{1}\cdot(\ar{2}\ars{1}{2}\vr{1}) \, , \spcc \Iv{2}{2}_{20}  =\vr{1}\cdot(\ars{1}{2}\ars{2}{2}\vr{1}) \, , \nn \\
\Iv{2}{2}_{21} &=& \vr{2}\cdot\ar{2}\vr{2} \, , \spcc
\Iv{2}{2}_{22} = \vr{2}\cdot\ars{2}{2}\vr{2} \, , \, , \nn \\
\Iv{2}{2}_{23}  &=& \vr{2}\cdot(\ar{1}\ar{2}\vr{2}) \, , \spcc \Iv{2}{2}_{24}  = \vr{2}\cdot(\ar{1}\ars{2}{2}\vr{2}) \, , \nn \\
\Iv{2}{2}_{25}  &=& \vr{2}\cdot(\ar{2}\ars{1}{2}\vr{2}) \, , \spcc \Iv{2}{2}_{26}  =\vr{2}\cdot(\ars{1}{2}\ars{2}{2}\vr{2}) \, , \nn \\
\Iv{2}{2}_{27} &=& (\vr{1}\cdot\vr{2})(\vr{1}\cdot \ar{1}\vr{2}) \, , \spcc
\Iv{2}{2}_{28} = (\vr{1}\cdot\vr{2})(\vr{1}\cdot \ar{2}\vr{2}) \, , \nn \\
\Iv{2}{2}_{29} &=&  (\vr{1}\cdot\vr{2})(\vr{1}\cdot \ar{1}\ar{2}\vr{2}) \, , \nn \\
\Iv{2}{2}_{30} &=& (\vr{1}\cdot\vr{2})(\vr{1}\cdot \ar{2}\ar{1}\vr{2}) \, , \nn \\
\Iv{2}{2}_{31} &=& (\vr{1}\cdot\vr{2})(\vr{1}\cdot \ars{1}{2}\ar{2}\vr{2}) \, , \nn \\
\Iv{2}{2}_{32} &=& (\vr{1}\cdot\vr{2})(\vr{1}\cdot \ar{2}\ars{1}{2}\vr{2}) \, , \nn \\
\Iv{2}{2}_{33} &=& (\vr{1}\cdot\vr{2})(\vr{1}\cdot \ar{1}\ars{2}{2}\vr{2}) \, , \nn \\
\Iv{2}{2}_{34} &=& (\vr{1}\cdot\vr{2})(\vr{1}\cdot \ars{2}{2}\ar{1}\vr{2}) \, , \nn \\
\Iv{2}{2}_{35} &=& (\vr{1}\cdot\vr{2})(\vr{1}\cdot \ars{2}{2}\ars{1}{2}\vr{2}) \, , \nn \\
\Iv{2}{2}_{36} &=& (\vr{1}\cdot\vr{2})(\vr{1}\cdot \ars{1}{2}\ars{2}{2}\vr{2}) \, , \nn \\
\Iv{2}{2}_{37} &=& (\vr{1}\cdot\vr{2})^2 \, . \ee 

From \rr{eig1} and \rr{ci-7a}, $\ld_i$ is expressed in terms of $\Iv{2}{2}_i$. From \rr{pr3a}, with  $r=1,2$, and \rr{ci-7b} we have
$6$ linear equations in $\cvs{r}_i$.  On solving $3$ linear equations for each $r$, we can express $\cvs{1}_i$ and $\cvs{2}_i$ explicitly in terms
of $\Iv{2}{2}_{\alpha} \, , \alpha=1,2 \ldots , 7$.  The invariants $\ca{2}_{ii}$ appear linearly in \rr{ci-7c}. On solving the $3$ linear
equations in \rr{ci-7c}, we can express $\ca{2}_{ii}$ explicitly in terms of 
$\Iv{2}{2}_{\alpha} \, , 
\alpha=1,2,3,8,9,10$. The invariants $\cas{2}_{ij} \, , i\ne j$ appear linearly in
the $3$ equations given by \rr{ci-7d}. On solving these equations we can express explicitly for $\cas{2}_{ij}$ in terms of
$\Iv{2}{2}_{\alpha} \, , 
\alpha=1,2,3,8,\ldots ,13$. The remaining classical invariants $\Iv{2}{2}_{\alpha} \, , \alpha=14,15, \ldots , 37$ in \rr{ic-7e} can be expressed
explicitly in terms of $\ld_i\, , \cv{r}_i\,  \ca{2}_{ij}$. Using the appropriate sign for
$\cv{r}$ and $\ca{2}_{ij} \, , i\ne j$, we can express the remaining classical invariants explicitly in terms of the $13$ independent
invariants $\Iv{2}{2}_\alpha \, , \alpha = 1,2 \ldots , 13$.

\subsection{Relations for general $m$ and $n$. At most $2m + 6n-3$ invariants are independent}
In this section we only construct relations between classical invariants for an isotropic function of the form
\be\label{ci-8}
W(\vr{1}\otimes\vr{1},\vr{2}\otimes\vr{2}, \ldots , \vr{m}\otimes\vr{m}, \ar{1},\ar{2}, \ldots , \ar{n}) \, . \ee 
Our intention is just to show relations, we shall not construct all the classical invariants in the general minimal integrity
basis, only the independent $2m+6n-3$ invariants, and they are:
\be\label{ci-8a} \Iv{m}{n}_1&=& I_1 \, , \spcc 
 \Iv{m}{n}_2 = I_2 \, , \spcc \Iv{m}{n}_3=  I_3 \, , \ee
\be\label{ci-8b}
\Iv{m}{n}_{2r+2} &=& \vr{r}\cdot\ar{1}\vr{r} = \sum_{i=1}^3 \cvs{r}_i\ld_i \, , \spcc \Iv{m}{n}_{2r+3} = \vr{r}\cdot\ars{1}{2} \vr{r} = \sum_{i=1}^3 \cvs{r}_i\ld_i^2 \, ,
\nn \\
  r&=&1,2, \ldots , m \, , \ee
\be\label{ci-8c}
\Iv{m}{n}_{2m+3+6s-5} &=& \tr\ar{s+1} = \sum_{i=1}^3 \ca{s+1}_{ii} \, , \nn \\
  \Iv{m}{n}_{2m+3+6s-4} &=& \tr(\ar{s+1}\ar{1}) = \sum_{i=1}^3 \ld_i \ca{s+1}_{ii} \, , \nn \\
 \Iv{m}{n}_{{2m+3+6s-3}} &=& \tr(\ar{s+1}\ars{1}{2}) = \sum_{i=1}^3 \ld_i^2\ca{s+1}_{ii}  \, , \ee
\be\label{ci-8d} \Iv{m}{n}_{{2m+3+6s-2}} &=& \tr(\ars{s+1}{2}) =  \cas{s+1}_{11} + \cas{s+1}_{22} + \cas{s+1}_{33} + \nn \\
&& 2(\cas{s+1}_{12} + \cas{s+1}_{13} + \cas{s+1}_{23}) \, ,  \nn \\
\Iv{m}{n}_{2m+3+6s-1} &=& \tr(\ars{2}{2}\ar{1}) =\ld_1(\cas{s+1}_{11} + \cas{s+1}_{12} + \cas{s+1}_{13}) + \nn \\ 
&& \ld_2(\cas{s+1}_{21} + \cas{s+1}_{22} + \cas{s+1}_{23}) + \nn \\
&& \ld_3(\cas{s+1}_{31}+ \cas{s+1}_{32} + \cas{s+1}_{33}) \, , \nn \\
\Iv{m}{n}_{2m+3+6s} &=& \tr(\ars{2}{2}\ars{1}{2}) = \ld_1^2(\cas{s+1}_{11} + \cas{s+1}_{12}+ \cas{s+1}_{13}) +  \nn \\
&& \ld_2^2(\cas{s+1}_{21} + \cas{s+1}_{22} + \cas{s+1}_{23}) + \nn \\
&& \ld_3^2(\cas{s+1}_{31} + \cas{s+1}_{32} + \cas{s+1}_{33}) \, , \nn \\
s&=& 1,2 \ldots n-1 \, , \spcc n > 1 \, . \ee

From \rr{eig1} and \rr{ci-8a}, $\ld_i$ is expressed in terms of $\Iv{m}{n}_i$. From \rr{pr3a}, with $r=1,2,\ldots , m$, and \rr{ci-8b} we have
$3m$ linear equations in $\cvs{r}_i$.  On solving the $3$ linear equations for each $r$, we can explicitly express, for all $\cvs{r}_i$, in terms
of $\Iv{m}{n}_{\alpha} \, , \alpha=1,2 \ldots , 2m+3$.  The invariants $\ca{s+1}_{ii} , s=1,2,\ldots , n-1$ appear linearly in \rr{ci-8c}. On solving the $3$ linear
equations in \rr{ci-8c} for each $s$, we can express $\ca{s+1}_{ii}$ explicitly in terms of 
$\Iv{m}{n}_{\alpha} \, , 
\alpha=1,2,3$ and $\alpha=2m+3+6s-5,2m+3+6s-4,2m+3+6s-3$. The invariants $\cas{s+1}_{ij} \, , i\ne j$ appear linearly in
the $3$ equations given by \rr{ci-8d}. On solving these equations for each $s$, we can  explicitly express $\cas{s+1}_{ij}$ in terms of
$\Iv{m}{n}_{\alpha} \, , 
\alpha=1,2,3,2m+4,\ldots ,2m+6n-3$. The remaining classical invariants $\Iv{m}{n}_{2m+6n-3+\alpha} \, , \alpha=1,2, \ldots$ can be expressed
explicitly in terms of $\ld_i\, , \cv{r}_i\,  \ca{s+1}_{ij}$. Using the appropriate sign for
$\cv{r}$ and $\ca{s+1}_{ij} \, , i\ne j$, we can express the remaining classical invariants explicitly in terms of the independent
invariants $\Iv{m}{n}_\alpha \, , \alpha = 1,2 \ldots , 2m+6n-3$.

Note that for $S(m,1)$, we have $2m+3$ independent invariants, which concur with the result of Shariff \cite{shariff16}. However, 
Shariff did not give classical invariant relations in his work \cite{shariff16}; the relations above for $S(m,1)$ supplement the results of
\cite{shariff16}. In the case of $S(0,n)$, we obtain $6n-3$ independent invariants; this agrees with the result of Shariff \cite{shariff18},
however, the relation forms  in \cite{shariff18} are different from the above.

\section*{Appendix A}
\def\theequation{A\arabic{equation}}
\setcounter{equation}{0}

In this Appendix we only give results for the case of $m=1$ and $n=2$ and when $\ld_1=\ld_2=\ld_3=\ld$.
Results when two of the eigenvalues $\ld_i$ are not-distinct are not given. Our main intention is to show that
at most $2m+6n-3$ invariants are independent and that when the  eigenvalues $\ar{s}$ are not distinct, the number of independent
invariants is less than $2m+6n-3$.

When $\ld_1=\ld_2=\ld_3=\ld$, the eigenvectors $\tu_i$ are arbitrary. We select $\tu_i$ to coincide with
the eigenvectors of $\ar{2}$. Hence, 
\be\label{app-1}
\ar{1} = \ld\tI \, , \spcc \ar{2}=\sum_{i=1}^3 \bar{\ld}_i \tu_i\otimes\tu_i \, . \ee
From Section \ref{sec-m1n2}, we have,
\be\label{app-6a} \Iv{1}{2}_1&=& 3\ld \, , \spcc 
\Iv{1}{2}_2 = 3\ld^2 \, , \spcc \Iv{1}{2}_3=  \ld^3 \, , \spcc
\Iv{1}{2}_4 =\ld \, , \spcc
\label{app-6b} \Iv{1}{2}_5 = \ld^2 \,  , \ee
\be\label{app-6c}
\Iv{1}{2}_6 =  \sum_{i=1}^3 \bar{\ld}_i \, , \spcc  \Iv{1}{2}_7 \tr(\ar{2}\ar{1}) = \ld\sum_{i=1}^3 \bar{\ld}_i \, , \spcc
 \Iv{1}{2}_8 = \ld^2\sum_{i=1}^3 \bar{\ld}_i  \, , \ee
\be\label{app-6d} \Iv{1}{2}_9 =  \sum_{i=1}^3 \bar{\ld}_i^2 \, , \spcc
\Iv{1}{2}_{10} = \ld\sum_{i=1}^3 \bar{\ld}_i^2 \, , \spcc 
\Iv{1}{2}_{11} = \ld^2\sum_{i=1}^3 \bar{\ld}_i^2 \, , \ee
\be\label{app-6e}
\Iv{1}{2}_{12} &=& \sum_{i=1}^3 \cvs{1}_i\bar{\ld}_i\, \, , \spcc
\Iv{1}{2}_{13} = \sum_{i=1}^3 \cvs{1}_i\bar{\ld}_i^2 \, , \spcc \Iv{1}{2}_{14}  = \sum_{i=1}^3 \bar{\ld}_i^3  \, , \nn \\
\Iv{1}{2}_{15}  &=& \ld\sum_{i=1}^3 \cvs{1}_i\bar{\ld}_i \, , \spcc \Iv{1}{2}_{16}  = \ld\sum_{i=1}^3 \cvs{1}_i\bar{\ld}_i^2\, , \nn \\
\Iv{1}{2}_{17}  &=& \ld^2\sum_{i=1}^3 \cvs{1}_i\bar{\ld}_i\, , \spcc \Iv{1}{2}_{18}  =\ld^2\sum_{i=1}^3 \cvs{1}_i\bar{\ld}_i^2\, . \ee

It is clear from \rr{app-6a} to \rr{app-6e} that only $6$ of the classical invariants are independent and we consider the invariants
\be\label{app-2}
\Iv{1}{2}_1 \, , \spcc \Iv{1}{2}_6 \, , \spcc \Iv{1}{2}_9 \, , \spcc \Iv{1}{2}_{12} \, , \spcc \Iv{1}{2}_{13} \, , \spcc
\Iv{1}{2}_{14} \, . \ee

We also have $6$ independent spectral invariants 
\be\label{app-3}
\ld \, , \spcc \ca{2}_{ii} =\bar{\ld}_i \, , \spcc \cv{1}_1 \, , \spcc \cv{1}_2 \, . \ee

The number of independent invariants is further reduced if, for example, $\bar{\ld}_1=\bar{\ld}_2=\bar{\ld}_3=\bar{\ld}$. In this case, we have,
\[ \Iv{1}{2}_6 = 3\bar{\ld} \, , \spcc \Iv{1}{2}_7 =3\ld\bar{\ld} \, , \spcc \Iv{1}{2}_8 =3\ld^2\bar{\ld} \, , \spcc
\Iv{1}{2}_9 =3\bar{\ld}^2 \, , \spcc \Iv{1}{2}_{10} =3\ld\bar{\ld}^2 \, , \]
\[ \Iv{1}{2}_{11}  =3\ld^2\bar{\ld}^2 \, , \spcc \Iv{1}{2}_{12} =\bar{\ld} \, , \spcc \Iv{1}{2}_{13} =\bar{\ld}^2 \, , \spcc
\Iv{1}{2}_{14} =\bar{\ld}^3 \, , \spcc \Iv{1}{2}_{15} =\ld\bar{\ld}  \, , \]
\be\label{app-4}
\Iv{1}{2}_{16} =\ld\bar{\ld}^2 \, , \spcc \Iv{1}{2}_{17} =\ld^2\bar{\ld}   \, , \spcc \Iv{1}{2}_{18} =\ld^2\bar{\ld}^2  \, . \ee

Hence, it is clear from \rr{app-4} that the classical invariants are independent of $\cv{1}_i$ and only $2$ of them are independent.
We can consider  the invariants 
\be\label{app-5} \Iv{1}{2}_1 \, , \spcc \Iv{1}{2}_6 \,  \ee
to be the independent invariants. However, the number of independent spectral invariants is $4$ and they are
\be\label{app-6}
\ld \, , \spcc \bar{\ld} \, , \spcc \cv{1}_1 \, , \spcc \cv{1}_2 \, . \ee

\section*{Appendix B}
\def\theequation{B\arabic{equation}}
\setcounter{equation}{0}
The solutions in the main body require the results
\be\label{apb1}
\left( \begin{array}{ccc}
                      1 &  1 & 1 \\
                      \ld_1 &  \ld_2 & \ld_3 \\
                      \ld_1^2 & \ld_2^2 &  \ld_3^2 \\
                      \end{array} \right)^{-1} = 
\left( \begin{array}{ccc}
                      \alpha_1\ld_2\ld_3  \spcc &  - \alpha_1(\ld_2+\ld_3) \spcc & \alpha_1 \\
                      \alpha_2\ld_1\ld_3 \spcc &  -\alpha_2(\ld_1+\ld_3) \spcc & \alpha_2 \\
                      \alpha_3\ld_1\ld_2 \spcc &  -\alpha_3(\ld_1+\ld_2) \spcc & \alpha_3 \\
                      \end{array} \right) \, 
\ee
and 
\be\label{apb3}
&&\left( \begin{array}{ccc}
                      2 &  2 & 2 \\
                      \ld_1+\ld_2 \,\, &  \ld_1+ \ld_3 \,\, & \ld_2+ \ld_3 \\
                      \ld_1^2 +\ld_2^2 & \ld_1^2 +\ld_3^2 &  \ld_2^2+\ld_3^2 \\
                      \end{array} \right)^{-1} \nn \\  
&& = \left( \begin{array}{ccc}
                      -\frd{1}{2}\alpha_3(\ld_1\ld_2+\ld_1\ld_3-\ld_3^2+\ld_2\ld_3) \spcc &  \alpha_3(\ld_1+\ld_2) \spcc & -\alpha_3 \\
                      -\frd{1}{2}\alpha_2(\ld_1\ld_2+\ld_1\ld_3-\ld_2^2+\ld_2\ld_3) \spcc &  \alpha_2(\ld_1+\ld_3) \spcc & -\alpha_2 \\
                      -\frd{1}{2}\alpha_1(\ld_1\ld_2+\ld_1\ld_3-\ld_1^2+\ld_2\ld_3) \spcc &  \alpha_1(\ld_2+\ld_3) \spcc & -\alpha_1 \\
                      \end{array} \right) \, ,
 \ee
where
\be\label{apb2}
\alpha_1 = \frd{1}{(\ld_1-\ld_2)(\ld_1-\ld_3)} \, , \spcc \alpha_2 = \frd{1}{(\ld_2-\ld_1)(\ld_2-\ld_3)} \, , \spcc
 \alpha_3 = \frd{1}{(\ld_3-\ld_1)(\ld_3-\ld_2)} \, . \ee

\end{document}